\documentstyle[pre,aps,eqsecnum,epsf]{revtex}

\newcommand{\ovl}[1]{\overline{#1}}
\newcommand{\beq}{\begin{equation}}
\newcommand{\eeq}{\end{equation}}
\newcommand{\beqn}{\begin{eqnarray}}
\newcommand{\eeqn}{\end{eqnarray}}
\newcommand{\dd}{\mbox{d}}

\newcommand{\xx}{{\bf x}}

\newcommand{\ee}{{\bf e}}

\newcommand{\Tb}{\overline{T}}
\newcommand{\EE}{{\cal E}}

\newcommand{\FF}{{\cal F}}

\newcommand{\RR}{{\cal R}}
\newcommand{\SS}{{\cal S}}
\newcommand{\nab}{\bbox \nabla}

\newcommand{\tB}{\tilde{B}}
\newcommand{\Bb}{\overline{B}}

\newcommand{\tx}{\tilde{x}}

\begin{document}

\draft 
\title{Aging and Non-Linear Glassy Dynamics in a Mean-Field Model}
\author{Fabrice{\sc Thalmann}$^{*,\dagger}$} 
\address{* LEPES-CNRS, Laboratoire associ\'e \`a l'UJF-Grenoble\\
{\it BP166X  25 avenue des Martyrs 38042 Grenoble Cedex} France\\
  and \\
$\dagger$ Department of Physics and Astronomy, University of Manchester\\
\textit{Oxford Road, Manchester M13 9PL} U.K.} 
\date{May~$5^{th}$, 2000}

\maketitle 

\begin{abstract}
  The mean-field approach of glassy dynamics successfully describes
  systems which are out-of-equilibrium in their low temperature
  phase. In some cases an aging behaviour is found, with no stationary
  regime ever reached. In the presence of dissipative forces however,
  the dynamics is indeed stationary, but still out-of-equilibrium, as
  inferred by a significant violation of the fluctuation dissipation
  theorem. The mean-field dynamics of a particle in a random but
  short-range correlated environment, offers the opportunity of
  observing both the aging and driven stationary regimes. Using a
  geometrical approach previously introduced by the author, we study
  here the relation between these two situations, in the pure
  relaxational limit, \textit{i.e.} the zero temperature case. In the
  stationary regime, the velocity ($v$)- force ($F$) characteristics
  is a power law $v\sim F^{4}$, while the characteristic times scale
  like powers of $v$, in agreement with an early proposal by
  Horner. The cross-over between the aging, linear-response regime and
  the non-linear stationary regime is smooth, and we propose a
  parametrisation of the correlation functions valid in both cases, by
  means of an ``effective time''. We conclude that aging and
  non-linear response are dual manifestations of a single
  out-of-equilibrium state, which might be a generic situation.
\end{abstract}

%

\pacs{PACS numbers~: 05.70.Ln, 64.70.Pf, 75.10.Nr, 83.50.Gd}

\section{Introduction}
\label{sec:intro}

Thermal equilibrium is the situation where all fast processes have
already taken place  while slow processes have not yet started
happening~\cite{Ma}. At the opposite, systems with slow dynamics are
characterised by a broad distribution of relaxation times, ranging
from the microscopic scale ($10^{-12}$ s) to the macroscopic one
(hours or days). For instance, glassy systems have an equilibration
time, either infinite, or much longer than the laboratory time
scale. These systems reveal their out-of-equilibrium state in
phenomena like \textit{aging} or \textit{non-linear response}.  

In these systems, the microscopic time scale is not the only relevant
one, and much slower processes also take place. The slow dynamics is
generally attributed to the presence of thermally activated barrier
crossing in the configuration space, but others mechanisms, such as
the so-called ``entropic barriers'' may also
contribute~\cite{barriers}. Glassy dynamics is observed when
the relaxation time $\tau$ becomes larger than the laboratory typical
time scale, as it is the case for supercooled
liquids~\cite{supercooled}. 

The aging behaviour of spin glasses has been thoroughly
investigated~\cite{aging-review}. The thermoremanent magnetization of
field cooled samples shows a strong waiting-time dependence (where the
waiting-time $t_w$ is the time interval between the temperature quench
and the measurement).  These systems have an \textit{a-priori}
infinite internal relaxation time, and the late stage of the
relaxation is instead controlled by the waiting-time itself. Moreover,
field cooled (waiting for $t_w$ and switching off the field) and zero
field cooled (waiting for $t_w$ and switching on the field) show a
remarkable complementarity of the magnetization
curves~\cite{VinHamOci,NorLunSan}. While out-of-equilibrium, as
indicated by its significant waiting-time dependence, the response of
the system is linear in the applied field, provided this one is weak
enough.

Glassy dynamics is also observed in the dissipative dynamics of
high-Tc superconductors. Supraconducting samples with quenched
disorder, at magnetic field and temperature large enough, offer a
significant resistance to a flowing dc current, due to the thermal
motion of the flux lines. A transition line is believed to separate an
ohmic regime (the vortex liquid) from a true superconducting state
(the vortex glass)~\cite{vortex-glass}. In the latter, and in
the limit of a vanishingly small current $j$, the dissipation occurs
by activation of ``bundles'' of flux lines over pinning energy
barriers. According to the scaling theory of the vortex glass, the
typical time needed for such a move, $\tau(j)$, diverges exponentially
fast as $j$ tend to zero~\cite{BlaGesFeiLarVin}. In this
situation, the response (the voltage) is a non-linear function of the
driving force (the current). The system is out-of-equilibrium, because
of a constant rate dissipation, but stationary, at variance with the
spin glass aging.  The relaxation time $\tau(j)$ which would be
infinite in the absence of driving force, is regularized by any small
but constant $j$, and inversely related to the magnitude of $j$.

Important and related issues are the supercooled liquids dynamics and
the rheology of soft glassy materials. In the first case, the
mode-coupling approach predicts an increase of the structural
relaxation time $\tau$ upon cooling~\cite{mode-coupling}. Aging have
been found during the early stage of molecular dynamics studies of the
Lennard-Jones fluid~\cite{Parisi,BarKob}.  On the other hand, a
constant shear rate flow seems to be able to change the value of
$\tau$, resulting in a shear rate dependent viscosity, {\it e.g.} a
non-linear response of the fluid~\cite{YamOnu:2}.  More generally,
this shear-thinning behaviour is well known in the context of
soft-matter rheology, and a phenomenological treatment of this
phenomenon based upon glassy dynamics has been
proposed~\cite{SolLeqHebCat,EvaCatSol}.  These are situations where
aging and non-linear response certainly coexist as manifestations of a
more general glassy dynamics.

Among the existing theoretical approaches on glassy systems, the
mean-field dynamics is a very promising one. It was already used to
suggest that the presence of dissipative forces generically prevents
the aging phenomenon~\cite{CugKurLeDPel,Kurchan}. In this framework, the
out-of-equilibrium character of a system is made precise by the
existence of a generalised fluctuation dissipation theorem, related in
turn to the entropy creation rate~\cite{CugDeaKur}.

A model of particular interest is the mean-field dynamics of a
particle with a quenched pinning potential. Isolated, the particle
presents an aging behaviour with a logarithmic growth of the time
correlation functions~\cite{FraMez,CugLeD}. In the presence
of a time-independent driving force, the dynamics is believed to be
stationary, with a power law dependence of the particle's velocity in
the applied force~\cite{Horner}.

Recently, the author presented a geometrical description of the aging
and linear response regime of this model, at zero
temperature~\cite{Myself:3}. This approach is extended, in this paper,
to the non-linear stationary regime. As a result, we find that aging,
linear response dynamics on the one hand, and stationary, non-linear
response dynamics on the other hand, are indeed dual manifestations of
a single out-of-equilibrium state.  The constant force is found to
interrupt efficiently the aging relaxational dynamics, and to control
the characteristic times, which in turn control the effective friction
coefficient in the stationary regime. The resulting velocity-force
characteristics is $v \propto F^{4}$, while the cross-over time
between aging and stationary regime is $t_F\propto F^{-3}$. These
predictions are confronted to the numerical integration of the
mean-field equations.  We finally suggest a scaling behaviour for the
correlation function of this model which, according to our numerical
findings, interpolate smoothly between the two different regimes,
demonstrating their common origin.

\section{Mean field equations and the Horner result}
\label{sec:two}

We focus on the zero temperature relaxational dynamics of a particle
in a quenched random gaussian potential~\cite{Myself:3}.  The particle
evolves in a $N$-dimensional space, under the simultaneous effect of a
pinning force $-\nab V$ and a constant force $\FF$, and the equation
of motion for the vector position $\xx(t)$ is~:
\begin{equation}
  \dot{{\bf x}}(t) = -{\nab} V({\bf x}(t)) + \FF.
  \label{eq:Langevin}
\end{equation}
The potential $V({\bf x})$ is a quenched disorder, chosen from a
gaussian distribution, with correlations (the overline stands for the
average over the quenched disorder)~:
\begin{equation}
  \ovl{V({\bf x})\cdot V({\bf x}')} = N\cdot
  \exp\left(-\frac{\|{\bf x}-{\bf x}'\|^2}{N} \right)\ ;\ 
  \ovl{V({\bf x})} = 0. 
  \label{eq:correlator}
\end{equation}
This form ensures a meaningful $N\to \infty$ limit, in which each
coordinate $\xx_i(t)$, or gradient component $\partial_i V({\bf x})$,
remains of order one, while the norms $\|{\bf x}(t)\|$, $\|{\nab} V\|$
scale like ~$N^{1/2}$. The force is directed along the direction~1.

The thermodynamic limit $N \to \infty$ is taken first, which makes the
zero temperature dynamics non trivial~\cite{KurLal}. In this
mean-field limit, the relaxation process is completely described by
the displacement $u$, the response function $r$ and the correlation
functions $b$ and $d$, $i\tilde{x}$ being the Martin-Siggia-Rose
auxiliary time~\cite{MarSigRos,DeD}.
\begin{eqnarray}
  u(t) & = & N^{-1/2}\ \ovl{ x_1(t) }; \\
  r(t,t') & = & N^{-1}\sum_{j=1}^{N} \ovl{ x_j(t)\cdot i\tx_j(t') }; \\
  b(t,t') & = & N^{-1}\sum_{j=2}^{N} \ovl{ (x_j(t)-x_j(t') )^2}; \\
  d(t,t') & = & N^{-1}\sum_{j=1}^{N} \ovl{ (x_j(t)-x_j(t') )^2};
  \nonumber\\ 
          & = & b(t,t') + [u(t)-u(t')]^2. \label{eq:motion:equations}
\end{eqnarray}
The Dyson equations for $r,b,d,u$ are a closed system of coupled
integro-differential equations, which contains the equations of
\cite{CugLeD,Horner} as a particular case~:
\begin{eqnarray}
  \partial_t r(t,t') & = & \delta(t-t') \nonumber\\
  & & \hspace{-1cm} -4\int_0^{t} \!\dd s\ \exp(-d(t,s))\ r(t,s)\
  [r(t,t') - r(s,t')];     \label{eq:Dyson-2temps:r}\\
   \partial_t b(t,t') & = & 
    (2T) -4\int_0^{t} \!\dd s\ \exp(-d(t,s)) \ [r(t,s) - r(t',s)]
   \nonumber\\ 
                      &   &  \hspace{-2cm}
   - 4\int_0^{t} \!\dd s\ \exp(-d(t,s))\
   r(t,s)\ [b(t,s) + b(t,t') -b(s,t')];  \label{eq:Dyson-2temps:b}\\
   \partial_t u(t)    & = & F - 4\int_0^{t} \!\dd s\ \exp(-d(t,s)) \
   r(t,s) \cdot [u(t) - u(s)].\nonumber\\
   & &  \label{eq:Dyson-2temps:u} 
\end{eqnarray}
The temperature term $(2T)$ is actually zero in our case.
It is also convenient to define the integrated response $\RR$ and the
energy $\EE$:
\begin{eqnarray}
  \RR(t,t') & = & \phantom{-2}\int_{t'}^{t}\!\dd s\ r(t,s);
    	\label{eq:def:int-response} \\
  \EE(t) & = & -2 \int_0^{t} \!\dd s\ \exp(-d(t,s)) r(t,s).
        \label{eq:def:energy}
\end{eqnarray}
In a seminal paper, Horner described the stationary state reached by
the system when driven with a finite force~$F$, in the case of short
range, power-law correlations~\cite{Horner}. We have found that the
system~(\ref{eq:Dyson-2temps:u}) does indeed lead to a stationary
situation, that we study in detail in this paper. One must mention
however that the stationary state reached by the particle depends on
how the system is prepared, in the same way as thermalised initial
conditions can prevent aging in the $p$-spin case~\cite{BarBurMez:2}.  The
stationary state is found only if the system is quenched from a high
enough temperature~\cite{Myself:5}. Results for a similar driven
system have also been recently published~\cite{BerBarKur}.

Let us summarise the main properties of the stationary solution found
by Horner~\cite{Horner}. The correlation functions are
time-translationally invariant (TTI)~: $r(t,t')=R(t-t')$,
$B(t,t')=B(t-t')$ while the displacement goes linearly with time
$u(t)=v\cdot t$. The system (\ref{eq:Dyson-2temps:u}) becomes a set of
non-causal equations to be solved self-consistently.  A non trivial
feature of this solution is the emergence of characteristic time
scales dependent on the velocity. With the notations of~\cite{Horner},
$t_p(v)$ is the characteristic time for breaking the fluctuation
dissipation theorem, while $t_a(v)$ controls the main ``$\alpha$''
relaxation of the correlation function $B(t)$.  These characteristic
times play a very similar role than the time scales $t_f$ and $t_b$
respectively, introduced in \cite{Myself:3}, and we identify
subsequently $t_f \equiv t_p$, $t_b \equiv t_a$. In the long time
regime, $t\sim t_b$:
\begin{equation}
  B(t) = q + \hat{B}\left( \frac{t}{t_b(v)} \right), 
   	\label{eq:scaling:stationary}
\end{equation}
with,
\begin{eqnarray}
  t_f(v) & \sim & v^{(\eta-1)\zeta}, \label{eq:tf-v} \\
  t_b(v) & \sim & v^{\eta-1},\; 0< \eta,\zeta < 1. \label{eq:tb-v} 
\end{eqnarray}
The exponents depend (in a complex way) on the
correlator~(\ref{eq:correlator})~\cite{Horner}.  $\hat{B}$ is a
scaling function discussed in appendix~\ref{app:two}, and $q$ is the
``plateau value'' of $B(t)$, equal to 0 in the zero temperature
limit. Meanwhile, the fluctuation dissipation theorem, obeyed for
$t\leq t_f$, is violated around $t\simeq t_f$, and becomes~:
\begin{equation}
  \dd B(t)/\dd t = 2\Tb R(t); \; t \gg t_f.
\end{equation}
The effective temperature $\Tb$ and the plateau value $q$ are
identical to those obtained in the aging case~\cite{CugLeD}.

The velocity-force characteristics is given by
(\ref{eq:Dyson-2temps:u}), and in the limit of small velocities,
\begin{equation}
  v \sim F / t_b(v). \label{eq:caracteristique:f-v}
\end{equation}
The time $t_b(v)$ plays the role of an effective friction coefficient,
controlled by the velocity. The $v-F$ characteristics is a power law.
\begin{equation}
  v \sim F^{1/\eta}.
\end{equation}
Let us mention for completeness the presence of a third time scale,
called $t'_a$ in~\cite{Horner}, defined by $B(t'_a) = v^2 {t'}_{a}^2$. As
we consider an exponential correlator, we have $\exp(-B(t)-v^2 t^2)
\equiv \exp(-B(t))\times \exp(-v^2 t^2)$ and in our case $t'_a\sim
v^{-1}$. We believe that apart from this point, the results of
\cite{Horner} all qualitatively apply to the exponential correlator,
and anyway $t'_a$ does not play a direct role in the dynamics of the
short range correlated models.

\section{Geometrical description of the driven stationary dynamics}
\label{sec:three}

In \cite{Myself:3} was proposed a geometrical description of the
relaxational dynamics of the particle. This approach makes use of a
comoving frame, defined by the eigendirections of the hessian matrix
$\nab\nab V(\xx)$ at the precise point $\xx(t)$ where the particle
stands. This frame is made of $N$ vectors $\{\ee_i\}$, each one
eigenvector of the hessian $\nab\nab V(\xx)$. The distribution of the
corresponding eigenvalues is a semi-circle of radius $4$, shifted
towards the positive values, and such that the lowest one is equal to
$-\SS$. Each eigenvector $\ee_i$ has an eigenvalue $\lambda_i-\SS$,
and the density of states of the $\lambda_i$ is~:
\begin{equation}
  \rho(\lambda)  = (8\pi)^{-1}
  \sqrt{\lambda\ (8-\lambda)}. 
        \label{eq:time-dep:spectrum}
\end{equation}
The quantity $\SS$ is positive, and depends linearly on the energy of
the system, \textit{e.g.} $\SS(t)= 4+2V(\xx(t))/N= 4+2\EE(t)$.  In the
aging case, $\SS(t)$ is a time-dependent function, while in the
stationary case, $\SS$ is constant.

One projects the instantaneous velocity~$\dot{\xx}$ is the above
frame such that:
\begin{equation}
  \dot{\xx} = \sum_{i=1}^{N} \gamma_i \ee_i
  \label{eq:definition-gamma-i}
\end{equation}
Because the spacing of the eigenvalues is of order $1/N$, the set of
$\lambda_i$ becomes dense, and one replaces the discrete sum over the
index $i$ by a continuous one, involving the semi-circular density of
eigenvalues $\rho(\lambda)$.
\begin{equation}
  \dot{\xx}^2 = \int_0^8 \dd\lambda\ \rho(\lambda)\ g(\lambda,t)
\end{equation}
The distribution $g(\lambda,t)$ represents the mean value of the
component $\xx_i^2$, locally averaged over the indices $i$ such that
$\lambda_i \simeq \lambda$. We have justified in~\cite{Myself:3} the
following self-similar form for $g(\lambda,t)$:
\begin{eqnarray}
  \xx_i^2 & \sim  & g(\lambda_i,t) \nonumber \\
  g(\lambda,t) &  = & \SS(t) \hat{G}\Big(\lambda/\SS(t)\Big),
	\label{eq:g-lambda}
	\end{eqnarray}
where $\lambda$ stands for any direction with a curvature of the
potential equal to $\lambda-\SS$~\cite{Myself:3}. The prefactor
$\SS$ in front of the distribution comes in fact from an assumption
about the value of the exponent $-\kappa$ governing the power law
decay of $\EE(t)$ and $\SS(t)\sim t^{-\kappa}$, which we believe to be
$-2/3$. This assumption is supported by our numerical results.

The characteristic times $t_f$ and $t_b$ are controlled by $\SS$, and
scale like~: 
\begin{eqnarray}
  t_f & \sim & \SS^{-1}; \nonumber \\
  t_b & \sim & \SS^{-3/2};
\end{eqnarray}
making the instantaneous velocity equal to~:
\begin{equation}
  \dot{u}(t) = F \cdot \SS^{3/2} = F/t_b(t). 
	\label{eq:inst-vel}
\end{equation}

When a constant force $F$ is applied, the dynamics changes from an
aging linear-response behaviour to a stationary regime~\cite{Myself:3}.
At short times, the displacement $u(t)$ is proportional to the
integrated response $F \cdot \RR(t,0)$, while at long times, it
becomes equal to $v\cdot t$.

\begin{figure}
    \epsfxsize 12cm  \centerline{
    \epsfbox{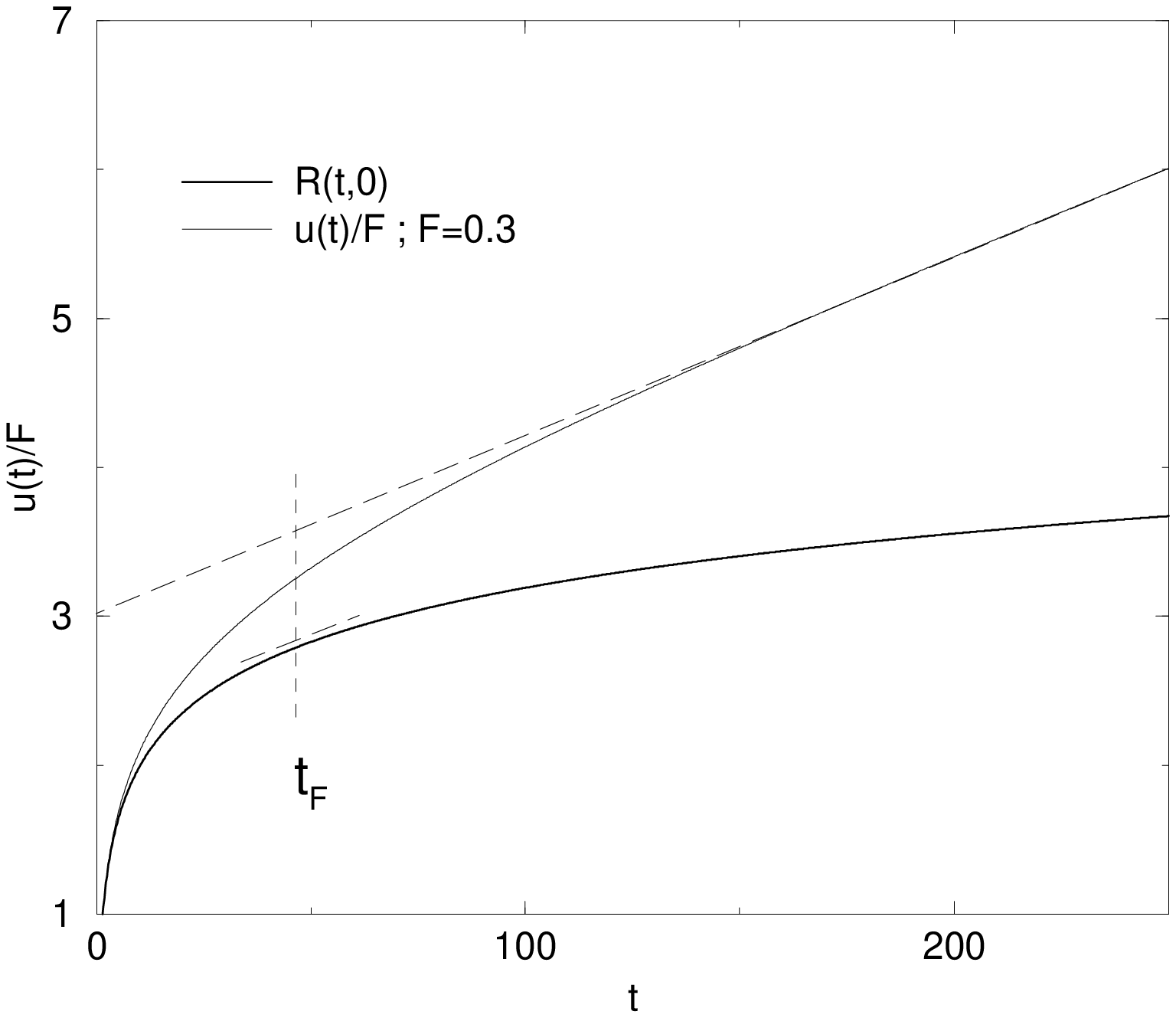} } 
    \caption{ }
    \label{fig:find-tf}
\end{figure}

One expects the stationary regime to take over the aging regime when
the dynamics is dominated by the external force $\FF$ rather than by
the gradient $-\nab V(\xx(t))$. This happens at a time $t_F$,
inversely related to the magnitude of the force.  Our numerical
definition of $t_F$, is the time where the slope of the asymptotic
curve $u(t)/F \simeq vt/F$ is equal to the slope of the (logarithmic)
integrated response $\RR(t,0)$, as shown on Figure~(\ref{fig:find-tf})
with $F=0.3$.  

\begin{figure}
    \epsfxsize 12cm  \centerline{
    \epsfbox{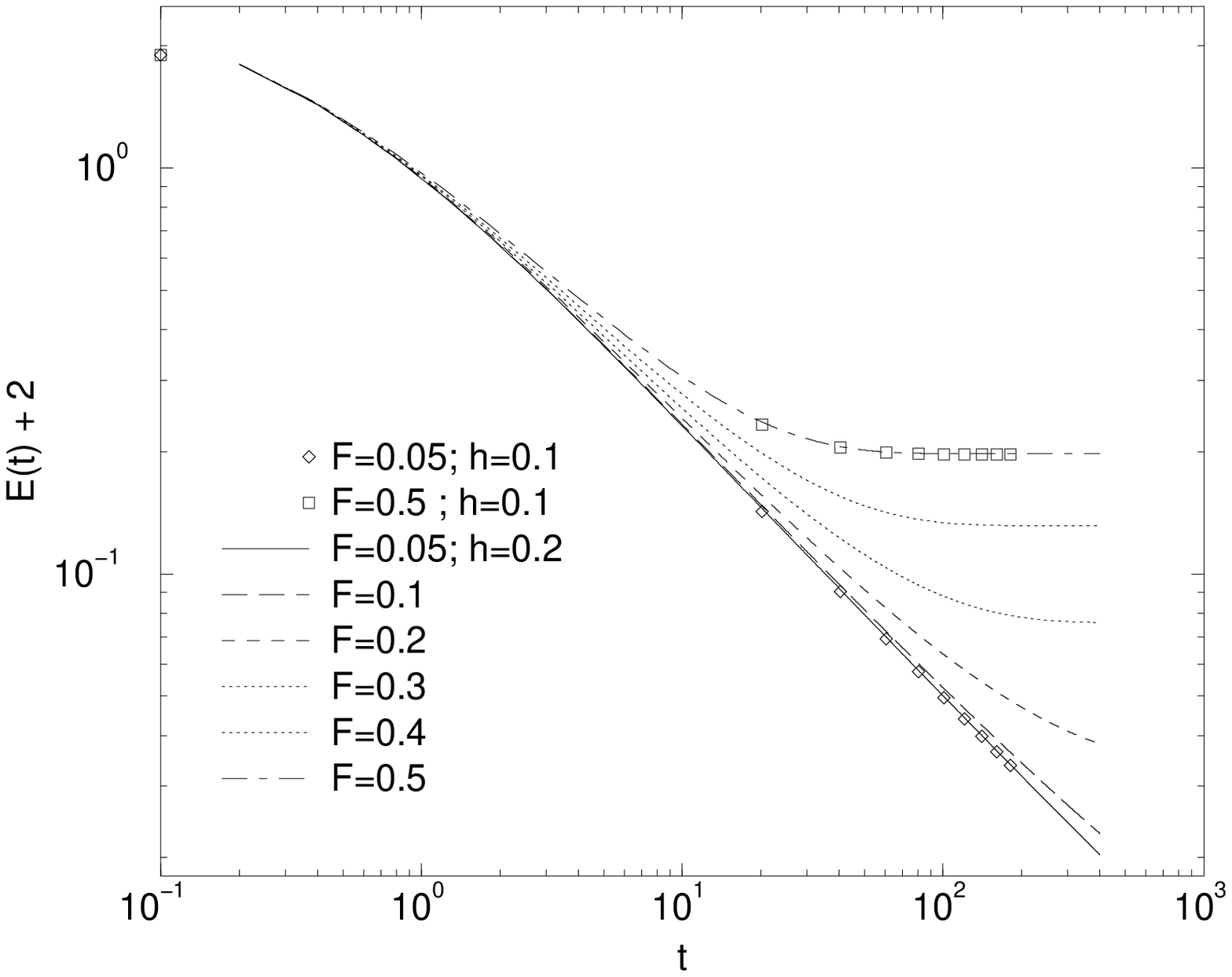} } 
    \caption{ }
    \label{fig:etf}
\end{figure}

In our situation, the dynamics is controlled by the value of $\SS$;
inverse friction and diffusion coefficients are both proportional to
$\SS^{3/2}$.  In the aging case, $\SS(t)$ tends to zero as a power law,
and the dynamics of the system is slower and slower.  A look at
Figure~(\ref{fig:etf}) however shows that in the presence of a force,
$\SS$ does not go to zero, but to a finite value $\SS(F)$, controlled
by $F$, and inversely related to the magnitude of $F$.  The same is
true for the energy $\EE(t)=-2+\SS(t)/2$, which stands higher than in
the absence of driving force.  Both diffusivity and mobility are kept
finite thanks to a non-zero driving force. What is needed is to
compute $\SS(F)$. For this purpose, one assumes that the self-similar
form (\ref{eq:g-lambda}) is still valid in the stationary regime. A
justification is provided in appendix~\ref{app:one}.

The zero temperature relaxation equation is:
\begin{equation}
  \dot{\xx}_i = -\partial_i V(\xx)+\FF_i,
\end{equation}
while the energy obeys~:
\begin{eqnarray}
  \dot{\EE}(t) & = & 1/N\sum_{i} \partial_i V(\xx(t)) \cdot
  \dot{\xx}_i(t),\nonumber\\
  \dot{\EE}(t) & = &  -\dot{\xx}^2(t)/N + F\cdot \dot{u}(t).
\end{eqnarray}

One uses now the distribution $g(\lambda,t)$ of the instantaneous velocity
components $\dot{\xx}_i^2$, the density of eigenvalues $\rho(\lambda)$
and finds~: 
\begin{equation}
  -\dot{\EE} (t) + F \cdot \dot{u}(t) = \int \dd \lambda\
   \rho(\lambda)\ g(\lambda,t). 
       \label{eq:general-relax}
\end{equation}
In the stationary regime, $\dot{\EE}(t)=0$ and $\dot{u}(t) = v$.
The equation (\ref{eq:general-relax}) reduces to a balance between the
mechanical power given by the force, and a kind of intrinsic dissipation
$(\dot{\xx}^2)$. 
\begin{equation}
 F\cdot v = \int \dd\lambda\ \rho(\lambda)\ g(\lambda).
\end{equation}
Assuming that $g$ is still equal to $\SS\hat{G}(\lambda/\SS)$ (cf
appendix~\ref{app:one}), one gets~:
\begin{equation}
  F\cdot v \sim \SS^{5/2}.
 	\label{eq:energy-balance}
\end{equation}
From (\ref{eq:energy-balance}) and (\ref{eq:inst-vel}), one finally finds $\SS$
as a function of the force,
\begin{equation}
  \SS  \sim  F^2,
	\label{eq:sf}
\end{equation}
the resulting velocity force characteristics,
\begin{equation}
  v \sim F^4,
	\label{eq:vf}
\end{equation}
and the force and velocity dependence of the time scales~:
\begin{eqnarray}
 	t_f & \sim & F^{-2} \sim v^{-1/2}; \\
	t_b & \sim & F^{-3} \sim v^{-3/4}.
\end{eqnarray}
These results are in full qualitative agreement with the findings of
Horner \cite{Horner}.
The main relaxation time $t_b$ does not scale as $v^{-1}$, as
could be expected from a simple dimensional analysis, but is shorter,
such that $\lim_{v\to 0} v\cdot t_b(v) =0$.

One determines the cross-over time $t_F$ by a matching argument. In
the linear response regime, $\SS(t)$ decreases as $t^{-2/3}$, as the
force acts only as a weak perturbation. The linear response breaks
down when the perturbation modifies the nature of the relaxation
itself. This happens when $\SS(t_F)$ reaches the order of magnitude of
its limit value $\SS(F)=F^2$ (equation~\ref{eq:sf}), leading to
$t_F^{-2/3} = F^2$, or:
\begin{equation}
  t_F = F^{-3}.
\end{equation}
Physically, this means that a typical coordinate $f_i$ of the force
$\FF$, along a downhill direction~$i$, is of the same order of
magnitude than the gradient of the potential $-\partial_i V$ , or the
instantaneous velocity $\dot{\xx}_i$~\cite{Myself:3}. From
equation~(\ref{eq:g-lambda}) and $f_i \simeq F$, one gets $F^2\simeq
f_i^2 \simeq \dot{\xx}_i^2 \simeq \SS$, in agreement
with~(\ref{eq:sf}).  

Let us mention that a qualitatively similar cross-over has been
observed in the simulated dynamics of a driven polymer, in the
presence of quenched disorder~\cite{Yoshino}.

\section{Numerical results}
\label{sec:four}

We present numerical results which support the findings of the
previous section.

\begin{figure}
    \epsfxsize 12cm  \centerline{
    \epsfbox{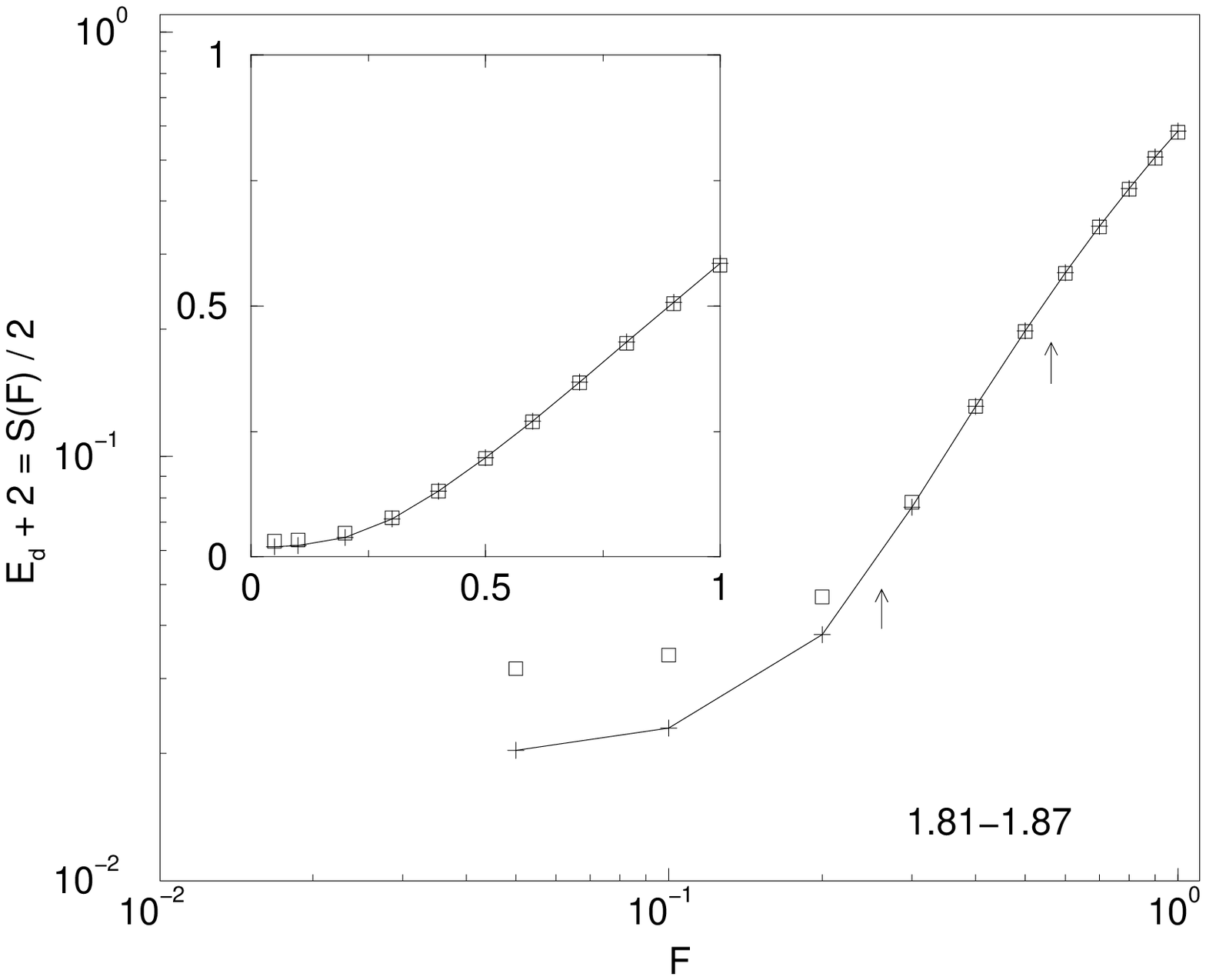} } 
    \caption{ }
    \label{fig:ef}
\end{figure}

Figure~(\ref{fig:ef}) shows $\SS(F)$ \textit{versus }$F$, in log
coordinates, and in regular coordinates (inset), for $F=0.05, 0.1,
0.2, 0.3, 0.4, 0.5$ and $F=0.6$.  The squares are the values obtained
with $t_{max}=200$ ($h=0.1$), and the full curve with $t_{max}=400$
($h=0.2$).  One sees that the three first values are not well
converged. If we except them, the overall shape of the curve is
concave (downward curvature). The slope of the tangent curve between
the arrows gives an exponent equal to $1.81$ ($h=0.1$) and $1.87$
($h=0.2$). Because the curve is concave, we believe that these values
are a lower bound for the real exponent, compatible with our
prediction $2$.

\begin{figure}
    \epsfxsize 12cm  \centerline{
    \epsfbox{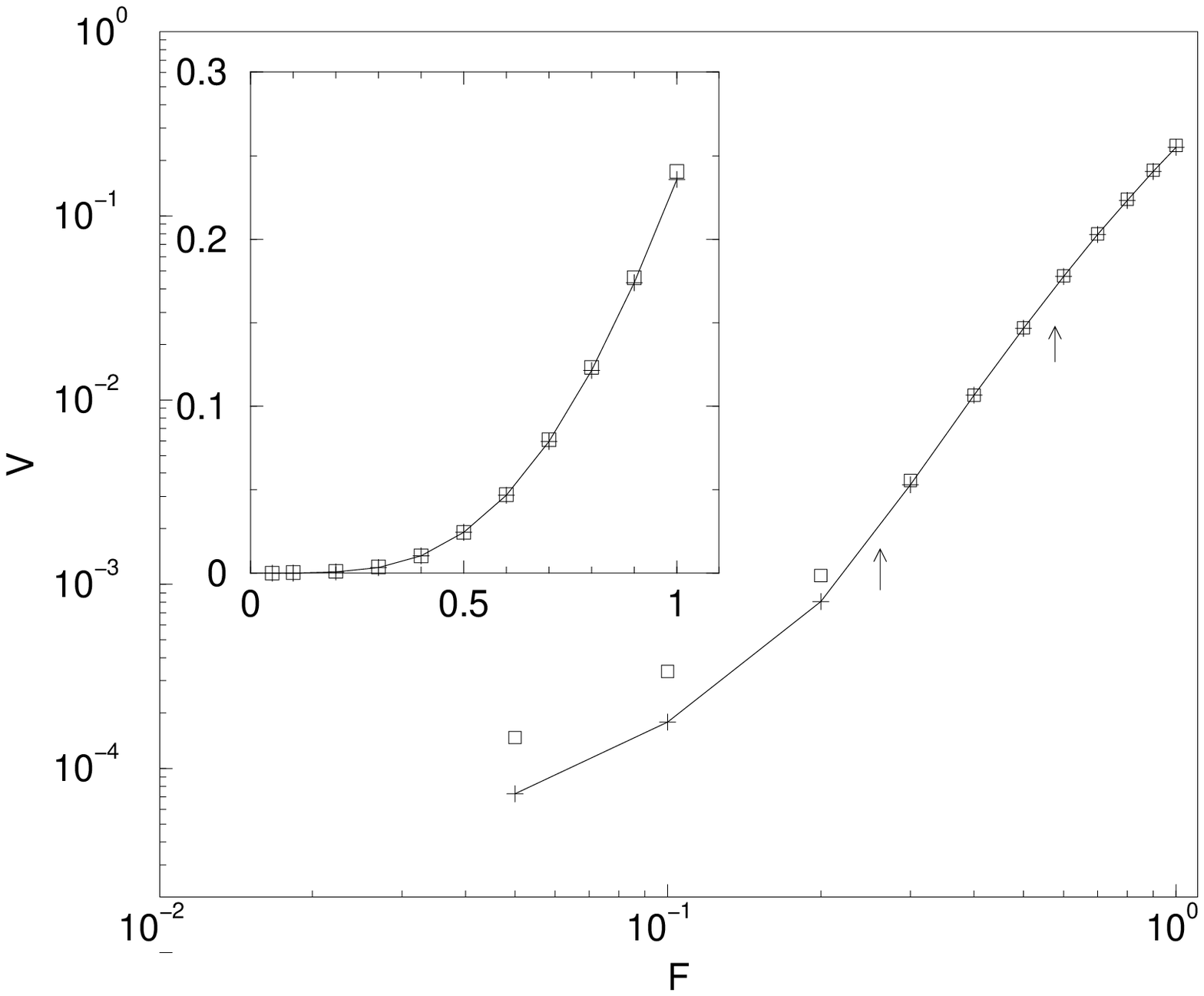} } 
    \caption{ }
    \label{fig:vf}
\end{figure}

Figure~(\ref{fig:vf}) shows $v(F)$ \textit{versus} $F$, in log
coordinates, and in regular coordinates (inset).  The squares are the
values obtained with $t_{max}=200$ ($h=0.1$), and the crosses with
$t_{max}=400$ ($h=0.2$).  As for Figure~(\ref{fig:ef}), the three
first values are not well converged. If we except them, the overall
shape of the curve is again concave. The slope of the tangent curve
between the arrows gives an exponent equal to $3.73$ ($h=0.1$) and
$3.82$ ($h=0.2$). Repeating the above argument, these values are a
lower bound for the real exponent, compatible with~4.

\begin{figure}
    \epsfxsize 12cm  \centerline{
    \epsfbox{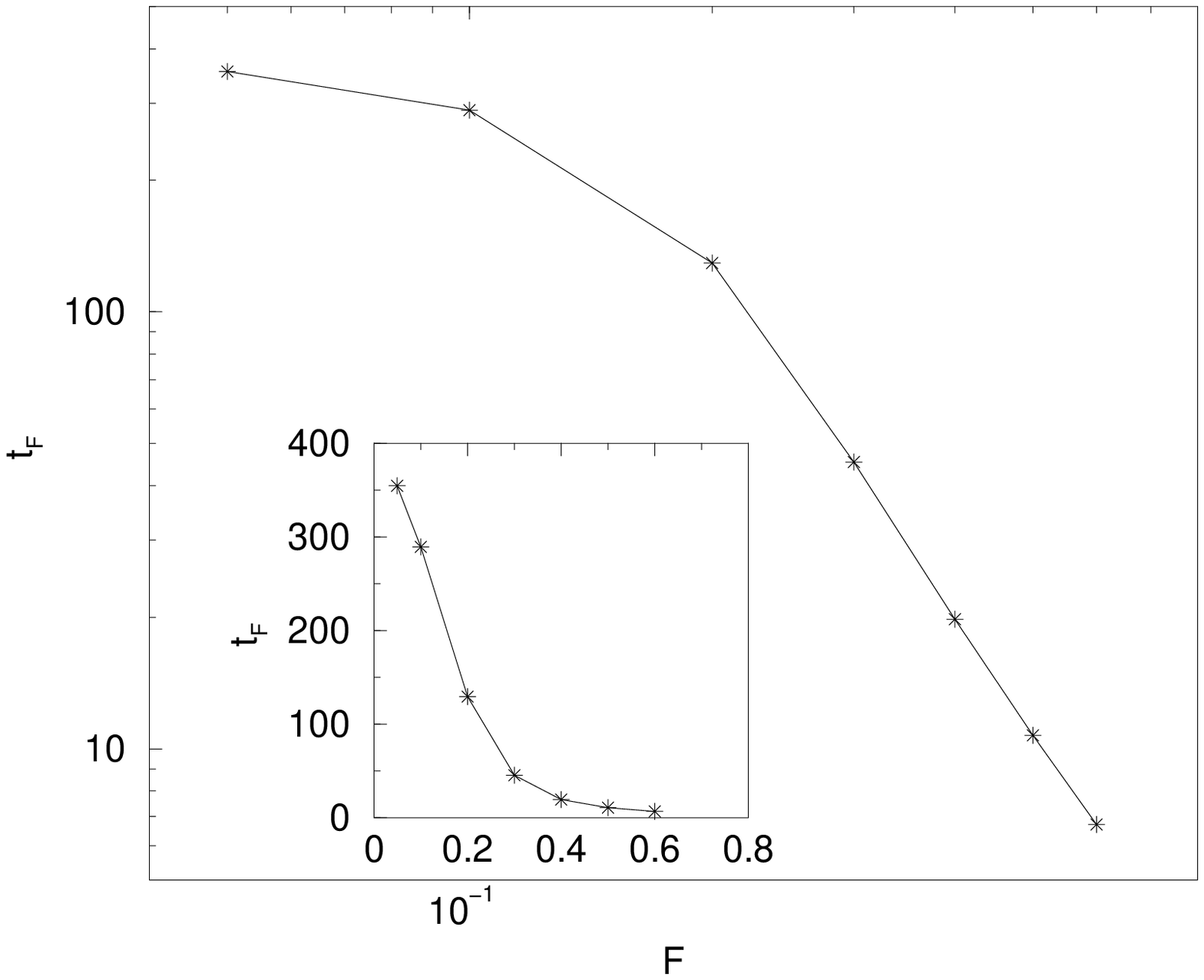} } 
    \caption{ }
    \label{fig:tco}
\end{figure}

A plot of $t_F$ \textit{vs} $F$ is reported on Figure~(\ref{fig:tco}).
Again the system has not reached its asymptotic regime as far as the
three first values $F\leq 0.2$ are concerned.  This can be checked by
looking at the first derivative $\dot{u}(t)$ which must be constant
when $t$ reaches the upper limit of the time window, here $t=400$. The
fitted value on the straight part of the graph, in logarithmic
coordinates, gives an exponent $-2.72$ instead of $-3$. Again the true
asymptotic limit $F \to 0$ is out of reach, due to our limited
computer facilities.

These numerical results are not good enough to prove the exactness of
the equations~(\ref{eq:sf}) and (\ref{eq:vf}). However they provide
lower bounds which constraint the exponents to be larger than $1.8$
for $\SS$, and larger than $3.8$ for $v$. On the other hand, if we
assume that we are close enough to the asymptotic regime where
equations (\ref{eq:sf}) and (\ref{eq:vf}) apply, one expects the real
exponents to be not too much different from the above numerical
values. In this respect, we think that the numerics is in agreement
with our findings.  As far as the cross-over time is concerned, the
numerical exponent is $-2.72$ instead of~$-3$. A larger time window
would certainly improve the agreement.


\section{A unified description of the out-of-equilibrium regimes}
\label{sec:five}

In the isolated aging regime, at zero temperature, the correlation
function obeys, as a particular case of
equation~(\ref{eq:scaling-aging-regime}) of appendix~\ref{app:two}, as
shown in~\cite{CugLeD}~:
\begin{equation}
  b(t,t') = \ln \left( \frac{h(t)}{h(t')} \right),
	\label{eq:scaling:long-time-aging}
\end{equation}
where the parametrisation function is related to the time-scale $t_b$
by~:
\begin{equation}
  t_b(t) = h(t)/h'(t)
\end{equation}

As the time scale $t_b$ is proportional to $\SS^{-3/2}$, we have~:
\begin{eqnarray}
  \frac{h(t)}{h(t')} &  = & \exp \left[ C \int_{t'}^{t} \dd s\
  \SS^{3/2} \right],\nonumber\\ 
  b(t,t') & = &  C \int_{t'}^{t} \dd s\ \SS^{3/2}.
	\label{eq:b-general}
\end{eqnarray}

On the other hand, equation (\ref{eq:inst-vel}) leads immediately to 
\begin{equation}
  \frac{u(t)-u(t')}{F} = C' \int_{t'}^{t} \dd s\ \SS^{3/2}.
	\label{eq:u-general}
\end{equation}

Now, one observes that the scaling form~(\ref{eq:scaling:stationary})
resemble to (\ref{eq:scaling:long-time-aging}),
(\ref{eq:scaling-aging-regime}), with $q=0$.  We prove in
appendix~\ref{app:two} that the scaling function of the aging
regime~\cite{CugLeD} and the driven regime~\cite{Horner} are indeed
equal, and thus $\hat{B}(x) = x$ in (\ref{eq:scaling:stationary})
(strictly speaking, $\hat{B}(x)$ is only proportional to $x$, but one
can choose $t_b$ such as $\hat{B}(x) = x$).  The equations
(\ref{eq:b-general}) and (\ref{eq:u-general}) make sense in the aging
regime as well as in the stationary regime.

The integral $\int_{t'}^{t}\dd s\ \SS^{3/2}$ is the \textit{effective
time} variable for the system, interpolating smoothly between $\ln(t)$
(aging, linear response regime) and $t/t_b(F)$ (stationary regime)
while $S^{-3/2}$ is an \textit{effective age}, growing like the
waiting time, in the aging regime, and bounded in the stationary
regime. Interestingly, a similar effective age has been used in the
context of the stick-slip motion in dry friction
experiments~\cite{HesBauPerCarCar}.

\begin{figure}
    \epsfxsize 12cm  \centerline{
    \epsfbox{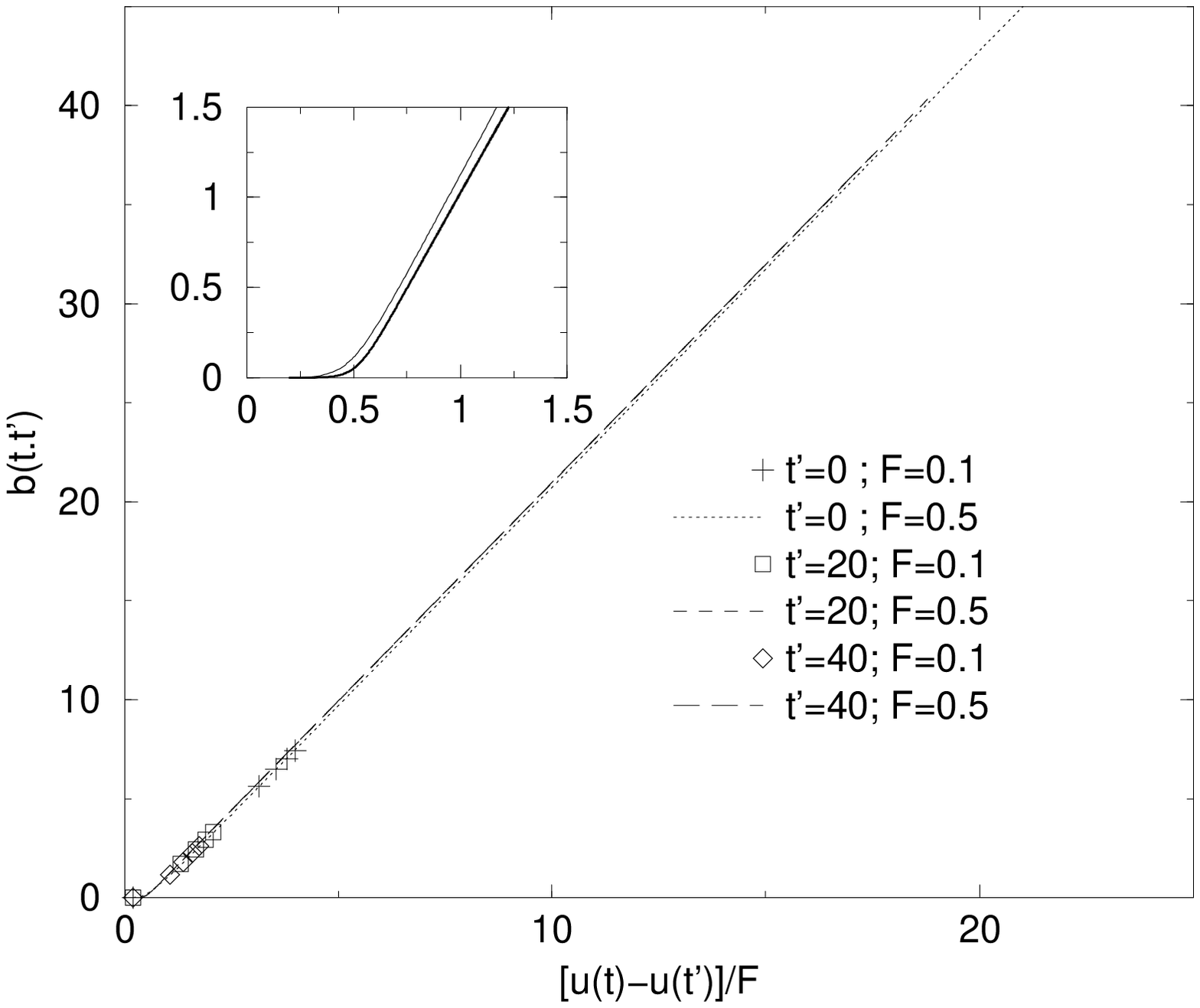} } 
    \caption{ }
    \label{fig:unif1}
\end{figure}

\begin{figure}
    \epsfxsize 12cm  \centerline{
    \epsfbox{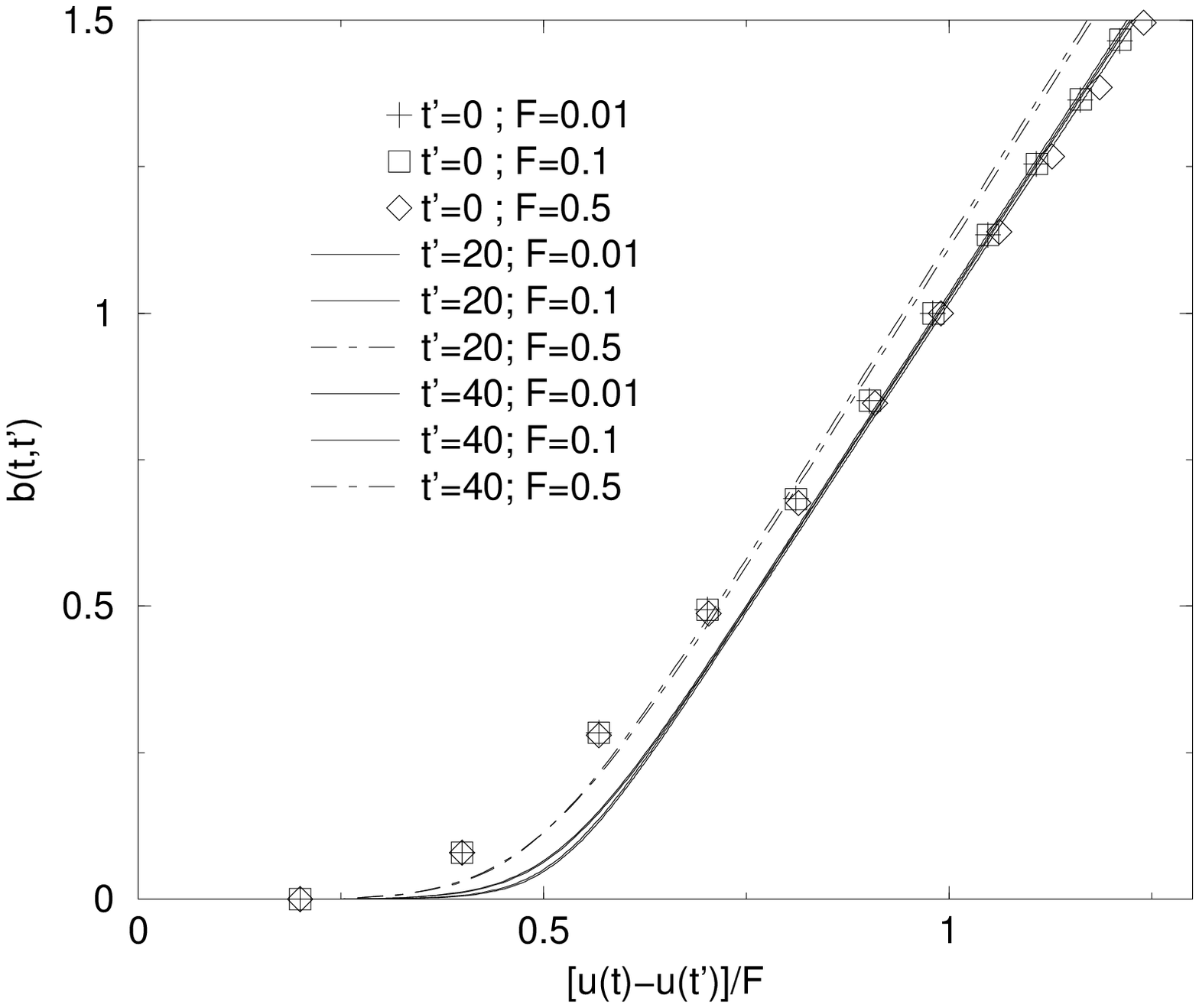} } 
    \caption{ }
    \label{fig:unif2}
\end{figure}

The prediction for (\ref{eq:b-general}) and (\ref{eq:u-general}) is
checked by plotting $b(t,t')$ vs $[u(t)-u(t')]/F$, shown on
Figures~(\ref{fig:unif1}) and~(\ref{fig:unif2}). One expects $b(t,t')$
and $[u(t)-u(t')]/F$ to be proportional, both in the aging and
stationary regimes, provided equations~(\ref{eq:b-general},
\ref{eq:u-general}) hold, which is the case for a time separation
$t-t'$ large enough.

On Figure~(\ref{fig:unif1}), $b(t,t')$ is plotted against
$[u(t)-u(t')]/F$ for $F=0.1$ (crosses, squares and diamonds) in the
linear response regime and for $F=0.5$ (continuous lines) in the
non-linear regime. The force is zero till $t=t'$, and then switched on;
 $t'$ takes the value 0, 20 and 40.  As far as $F=0.5$ is
concerned, the transition from linear to non-linear regime is not
visible on this curve, and in any case very smooth.  The slope of the
curve defines the effective temperature $2\Tb$, equal to the ratio
$C/C'$ in equations (\ref{eq:b-general}) and (\ref{eq:u-general}).
The effective temperature thus makes sense in both linear and non-linear
regimes.  

As the force is switched on at $t'$, there is a short-time ``elastic''
displacement. This is how the directions with a positive curvature
respond to the new static constraint, and this corresponds to the
short horizontal step at the origin, seen on Figures~(\ref{fig:unif1})
(inset) and~(\ref{fig:unif2}).  The finite slope part of the curve
corresponds to the slow wandering motion of the particle in the energy
landscape, in the regime where equations (\ref{eq:b-general}) and
(\ref{eq:u-general}) apply. Thus, we conclude that
Figure~(\ref{fig:unif1}) support the proportionality of
$u(t)-u(t')$ and $b(t,t')$, once the short time regime has been taken
into account.

A close look near the origin of the graph (inset of
Figure~(\ref{fig:unif1}) shows that the $F=0.5$ curve is slightly
shifted from $F=0.1$, but parallel to it. This shift goes rapidly to
zero as $F\to 0$. The shift is presumably there because $0.5$ is
already a large value of the force, leading to a departure from the
ideal curve corresponding to $F \ll 1$.

Figure~(\ref{fig:unif2}) is the same as Figure~(\ref{fig:unif1}) for
$F=0.01$, $F=0.1$ and $F=0.5$, for three values of $t'$, 0, 20 and
40, and gives additional details on the short time response of the
particle. Again, the horizontal part of the curves corresponds to the
short-time displacement (``elastic'' or reversible) while the
finite slope regime corresponds to the slow motion in the energy
landscape (``plastic'' or irreversible).

\section{Conclusion}
\label{sec:conclusion}

In this paper, we have proposed a consistent picture for the
stationary driven dynamics, in the mean field approximation and zero
temperature limit, of a particle in a quenched, exponentially
correlated, random potential.

The velocity $(v)$- force $(F)$ relation is a power law $v=F^4$, while
the main relaxation time scales as $t_b \simeq v^{-3/4}$. The product
$v\cdot t_b$ tends to zero as $v$ vanishes. These findings are
consistent with earlier work~\cite{Horner}. The driving force is found
to generate a relaxation time smaller than the ``dimensional'' time
scale $v^{-1}$, which is probably a generic feature of the mean-field
short-range correlated potentials.

If the force $F$ is small enough, a linear response around the aging
regime is found, up to a time $t_F$, scaling as~$F^{-3}$.  A plot of
the displacement $(u(t)-u(t'))/F$ \textit{vs} the correlation
$b(t,t')$ shows no sign of discontinuity, when the linear response
regime is replaced with the non-linear stationary regime. We interpret
it by saying that, when a small force is applied, the dynamical
properties of the system (mobility, diffusivity) are controlled by the
\textit{effective age} $\SS^{-3/2}$. The quantity $\SS^{3/2}$ is
proportional to the number of negative eigenvalues in the spectrum of
the hessian of the hamiltonian. The effective age is proportional to
the waiting time in the aging regime, and finite in the stationary
case.

The \textit{effective time} $\int^{t} \SS^{3/2} \dd s $, closely
related to the correlation function $b(t,t_0)$, grows logarithmically
with $t$ in the aging regime, and linearly with $t$ in the stationary
regime.  The effective temperature $\Tb$ generalising the fluctuation
dissipation theorem, remains unchanged in the non-linear
regime. However, the geometrical meaning of $\Tb$, if any, is still
unknown.

Future work will determine to what extent are the present features
generic from other short range correlated models, and finite
dimensional models. Even though such a power law dependence of the
characteristic times in the driving force is not observed in
realistic systems, the qualitative behaviour presented in this study
--cross-over between linear to non-linear regime, coexistence of aging
and non-linear stationary dynamics--, could indeed be a very generic
situation.

\paragraph*{Acknowledgements} I especially thank L.Cugliandolo and
J.Kurchan for having lent me their numerical code, and S.Scheidl, J.P
Bouchaud, J.Kurchan, M.M\'ezard and A.Cavagna for discussions on this
field.  I thank D.Feinberg for suggestions and criticisms about the
manuscript.  I warmly thank the hospitality of the Department of
Physics, IISc, Bangalore, where a part of the writing has been done.

\appendix
\section{The energy balance}
\label{app:one}

Let $\gamma_i(t)$ be the coordinates of the instantaneous velocity
$\dot{\xx}(t)$ in the comoving frame $\{ \ee_i(t) \}$
(\ref{eq:definition-gamma-i}). When $\FF=0$, this definition is
equivalent to say that $\gamma_i$ is the coordinate of $-\nab V$. Using
the local average defined in~\cite{Myself:3}, one finds:
\begin{eqnarray}
  \|\dot{\xx}\|^2(t) & = & \sum_{i} \dot{\xx}_i^2(t) = \sum_{i}
  \gamma_i^2(t), \nonumber\\  
  & = & N \int \dd \lambda\ \rho(\lambda)\ g(\lambda,t).   
\end{eqnarray}
The derivative of $\|\dot{\xx}\|^2(t)$ reads~:
\begin{eqnarray}
  \partial_t \|\dot{ \xx }\|^2(t) & = &  
  \sum_j^{N} \partial_t (-\partial_j V + \FF_j)^2 \nonumber\\ 
  & = & -2 \sum_{jk} \partial_{jk} V \cdot \dot{\xx}_j \cdot \dot{\xx}_k
  \nonumber\\ 
  & = & -2N \int \dd \lambda\ \rho(\lambda)\ (\lambda - \SS)\cdot
  g(\lambda,t).
\end{eqnarray}

We deduce that, in the stationary situation, for all $\SS$,
\begin{equation} 
  \int \dd\lambda\ \rho(\lambda)\ \lambda\ g(\lambda) = \SS \int
  \dd\lambda\ \rho(\lambda)\ g(\lambda), 
\end{equation} 
which is in favour of a scaling form $g(\lambda)= \Gamma \hat{G}
(\lambda/\SS)$.

As $\partial_i V(\xx(t)) = -\dot{\xx}_i(t) +\FF_i$, the
equation for $\dot{\EE}(t)$ is~:
\begin{eqnarray}
\dot{\EE}(t) & = & N^{-1} \sum_j \partial_j V \cdot \dot{\xx}_j
\nonumber\\
  & = & N^{-1} \sum_j \{ -\dot{\xx}_j^2(t) + \FF_j\cdot \dot{\xx}_j(t) \} 
\nonumber
\end{eqnarray}
The product $N^{-1} \sum_j \FF_j\cdot \dot{\xx}_j$ is by construction
equal to $F\cdot \dot{u}(t)$. Thus, (this is
equation~\ref{eq:general-relax}): 
\begin{equation}
  \dot{\EE}(t) = -\int \dd\lambda\ \rho(\lambda)\ g(\lambda,t) +
  F \cdot \dot{u}(t).
\end{equation}

The energy balance (\ref{eq:energy-balance}), and the factorised form
of $g(\lambda)$ imply in the stationary regime~:
\begin{equation}  
  \Gamma\cdot \SS^{3/2}  \propto F^2 \cdot \SS^{3/2}
\end{equation}
However the relation between $\SS$ and $\Gamma$ remains undetermined
by the present argument.  For the sake of simplicity, we can suppose
that the equality $\SS = \Gamma$, true in the aging regime, remains
true in the stationary regime.  This assumption is in fact equivalent
to a matching argument, when the distribution
$g(\lambda,t)=\SS(t)\hat{G}(\lambda/\SS(t))$, crosses over the
distribution $g(\lambda)= F^2 \hat{G}(\lambda/\SS(F))$ around
$t=t_F$. The matching of $g(\lambda,t)$ and $g(\lambda)$ leads to the
identification $\SS(F)=F^2$. One cannot rule out, rigorously, more
complicated behaviours, which could lead to a different velocity-force
characteristics. The assumption $\Gamma=\SS$ is just the most
natural one.

\section{The scaling form of the correlation function}
\label{app:two}

In the isolated situation, the correlation function in the aging
regime reads, for any finite temperature~$T$~\cite{CugLeD,CugKur:2}:
\beq
  b(t,t') = q + \tilde{B}\left[ \ln\left(\frac{h(t)}{h(t')}\right)
  \right] 
  \label{eq:scaling-aging-regime}
\eeq
An general equation for $\tilde{B}(u)$ is obtained in reference~\cite{CugLeD}
(equation 6.22, with the opposite sign convention for $f$), and reads:
\beqn
0 & = & \tB(u)f''(q) -f'(q+\tB(u)) +f'(q) \nonumber\\
  &   & +\frac{2\chi q}{T}f''(q) \int_{0}^{u} \dd u'\ \tB'(u') f''(q+\tB(u))
  \tilde{B}(u-u') \label{eq:tilde-B}
\eeqn
whose solution is $\tB(u) = C^{st}\times u $, leading
to~(\ref{eq:scaling:long-time-aging}), with $q=0$. The function
$f=\exp(-x)$ stands for the correlator~(\ref{eq:correlator}), $T$ for
the temperature, $q$ for plateau value of the correlation function
$b$, and $\chi$ for the fluctuation dissipation violation parameter
(see equation~\ref{eq:FDT-violation} below).

On the other hand, in the stationary regime, the equation for
$b(t,t')=B(t-t')=q+\Bb(t-t')$ and $R(t-t')=r(t,t')$ is (equation (2.9)
in reference~\cite{Horner}, again with the opposite sign convention
for $f$):
\beqn
\partial_t B(t) & = & 2T -\left( \int_0^{\infty} \dd s\
4f''(B(s)+v^{2}s^{2}) R(s) \right)\cdot B(t) + \int_0^{t} \dd s\ 
4f''(B(s)+v^2s^2) R(s) B(t-s) \nonumber\\
 & & + \int_0^{\infty} \dd s\ \Big\lbrace\Big( 4 f'(B(t+s)+v^2(t+s)^2)
 - 4f'(B(s)+v^{2}s^{2})\Big) R(s)  \nonumber\\
 & &  +\Big( 4f''(B(t+s) +v^2(t+s)^2) R(t+s)
 -4f''(B(s)+v^2s^2)R(s) \Big) B(s) \Big\rbrace
  \label{eq:intermediate-1} 
\eeqn
One knows that the main relaxation scale $t_b$ is much smaller than
$v^{-1}$, and asymptotically, $\lim_{t\to 0} v\cdot t_b = 0$. The
above integrals can be safely cut beyond a cut-off $\Lambda$ such that
$t_b \ll \Lambda \ll v^{-1}$. The contributions
$\int_{\Lambda}^{\infty}$ are negligible because the relaxation
of $B(t)$  has already taken place, while in the integrals
$\int_{0}^{\Lambda}$, the term $v^2 s^2$ can be neglected compared
with $B(s)$ in the argument of the correlators $f'$ and $f''$.

One introduces the quasi fluctuation dissipation parameter $X$, defined
by:
\beq
   - X(\Bb(t)) \cdot\dd B(t) /\dd t = R(t). \label{eq:FDT-violation}
\eeq
$X$ is equal to its equilibrium value $-1/2T$ if $\Bb<0$ and to $\chi$
if $\Bb>0$. Equation (\ref{eq:intermediate-1}) becomes:
\beqn
\partial_t B(t) & = & 2T + \left( \int_{0}^{\Lambda} \dd s\
4f''(q+\Bb(s)) X(\Bb(s)) \dd \Bb(s)/\dd s\ \right) B(t) 
- \int_{0}^{t} 4f''(\Bb(s)) X(\Bb(s)) \dd \Bb(s)/\dd s\ \cdot
B(t-s)\nonumber\\
& & - \int_0^{\Lambda} \dd s\ \Big\lbrace \Big( 4f'(q +\Bb(t+s))
-4f'(q+\Bb(s))\Big) X(\Bb(s)) \dd \Bb(s)/\dd s \nonumber\\
& & +\Big(4f''(q+\Bb(t+s)) X(\Bb(t+s)) \dd \Bb(t+s)/\dd s -4f''(q+\Bb(s))
X(\Bb(s)) \dd \Bb(s)/\dd s \Big) \Big\rbrace B(t)
\eeqn
Each integral $\int_a^b $ has to be split to take into account the
short time quasi-equilibrium regime and the long time regime. 
As the time scale $t_f$ separates these two regimes, one writes
$\int_a^b = \int_a^{a+t_f} + \int_{a+t_f}^{b-t_f} + \int_{b-t_f}^b$. 
The parameter $X$ is then set to $-1/2T$ or $\chi$ accordingly, and
most of the integrals can be reduced to boundary terms.
One neglects the time derivative $\partial_t B(t)$ in the asymptotic
long-time regime, and the result is:
\beqn
0 & = & 2T +\left( 4\chi\int_{t_f\simeq 0}^{\Lambda\simeq\infty}\!
f''(q+\Bb) \dd \Bb \right) (q+\Bb(t)) \nonumber \\
 & & - 4q \left( \chi + \frac{1}{2T}\right)\times\Big( f'(q)-f'(q+\Bb(t))
 \Big) \nonumber\\
 & & - 4\chi \int_{t_f\simeq 0}^{t} \dd s\ \dd \Bb(s)/\dd s\ f''(q+\Bb(s))
  (q+\Bb(t-s)) \label{eq:hat-B}
\eeqn
By using $\lim_{t\to\infty} \Bb(t) = \infty$, $q^2 f''(q)=T^2$ and
$-4\chi\int f''(\Bb)\dd \Bb = 2T/q$, the equation (\ref{eq:hat-B}) for
$\Bb$ coincides exactly with (\ref{eq:tilde-B}). As the equation
(\ref{eq:hat-B}) is invariant upon time dilatations, $\hat{B}(u) =
\Bb(t/t_b)$ is a solution of (\ref{eq:tilde-B}) and without loss of
generality, one has:
\beq
  \hat{B}(u) = \tilde{B}(u) = u,
\eeq
which is the announced result.



\clearpage
\centerline{\Large Captions}
\bigskip

FIGURE~1. Determination of the cross-over time $t_F$, defined as the
time where the slope of the integrated response $\RR(t,0)$ is equal to
the velocity $1/F \lim_{t\to\infty}\ \dot{u}(t)$. 

FIGURE~2. Family of curves $\EE(t)+2$ for increasing forces, ranging
from $F = 0.05$ to 0.5. The limit value $\lim_{t\to \infty} \EE(t)+2$
is a monotonically increasing function of $F$, equal to $\SS(F)/2$.
The effective mobility and diffusivity are directly related to
$\SS(F)$. The system stays above the marginal states, in a region
with a finite extensive number of downhill directions.

FIGURE~3 The parameter $\SS$ as a function of the force, for $F=0.1,
0.2, 0.3, 0.4, 0.5$ and $0.6$, in log coordinates, and normal
coordinates (inset). The boxes stand for a run, up to a time $t=200$
while the straight line corresponds to $t=400$. Whenever the boxes
differ from the line, the value is not converged. See text for
details. 
 
FIGURE~4. The velocity $v$ as a function of the force $F$, in log
coordinates, and normal coordinates (inset). Same remark as for
Figure~(4).

FIGURE~5. The time $t_F$ as a function of the force $F$ in log
coordinates. Inset~: $t_F$ as a function of $F$. The three first
values are not accurate ($t_F$ larger than our maximum time).  The
fitted exponent of the straight part is $-2.72$ instead of $3$;
$-2.72$ is a lower bound for the real value.

FIGURE~6. The correlation $b(t,t')$ \textit{vs} the displacement
$[u(t)-u(t')]/F$, for $F=0.1$ and $0.5$. The force is switched on at
$t'$, successively equal to $0, 20$ and $40$. Inset: the short-time
behaviour. See text for details.
 
FIGURE~7. Same as Figure~(6), with $F=0.01$, $F=0.1$ and $F=0.5$.

\clearpage


\end{document}